\documentclass[twocolumn,pra,superscriptaddress,noshowpacs,epsf]{revtex4}

\usepackage{graphicx}
\usepackage{amssymb}
\usepackage{comment}
\usepackage{amsmath}
\usepackage{color}

\begin{document}
\title{Probing the non-locality of Majorana fermions via quantum correlations}
\date{\today}
\author{Jun Li}
\affiliation{Beijing Computational Science Research Center, Beijing
100084, China}
\affiliation{College of Physics and Electronic Engineering, Dezhou University, Dezhou 253023, China}
\author{Ting Yu}
\affiliation{Center for Controlled Quantum Systems and
Department of Physics and Engineering Physics, Stevens Institute of
Technology, Hoboken, New Jersey 07030, USA}
\author{Hai-Qing Lin}
\affiliation{Beijing Computational Science Research Center, Beijing
100084, China}
\author{J. Q. You}
\affiliation{Beijing Computational Science Research Center, Beijing
100084, China}

\begin{abstract}
Majorana fermions (MFs) are exotic particles that are their own
anti-particles. Recently, the search for the MFs occurring as
quasi-particle excitations in solid-state systems has attracted widespread
interest, because of their fundamental importance in fundamental physics
and potential applications in topological quantum computation based on
solid-state devices. Here we study the
quantum correlations between two spatially separate quantum dots induced by a
pair of MFs emerging at the two ends of a semiconductor nanowire, in order to develop a new method for probing the MFs.
We find that without the tunnel coupling between these paired MFs, quantum
entanglement cannot be induced from an unentangled (i.e., product) state, but quantum discord is observed due to the
intrinsic nonlocal correlations of the paired MFs.
This finding reveals that quantum discord can indeed demonstrate the
intrinsic non-locality of the MFs formed in the nanowire. Also, quantum discord can be employed to discriminate the MFs from the regular fermions. Furthermore, we propose an experimental setup to measure the onset of
quantum discord due to the nonlocal correlations. Our approach provides a new, and experimentally accessible, method to study the Majorana bound states by probing their intrinsic non-locality signature.
\end{abstract}

\maketitle

Quantum entanglement is a quantum correlation that has no classical analog. Dynamical entanglement has exhibited some exotic properties such as the early-stage decoherence~\cite{decoherence1}
and revival~\cite{death}. Recently, it was found that
there exists another type of nonclassical correlation coined as quantum
discord~\cite{discordVedral,discordPRL}. A remarkable feature of
the quantum discord is that it can characterize some
fundamental processes of physics, including the quantum phase
transitions~\cite{phasetransition2}
and the discrimination of quantum and classical Maxwell's
demons~\cite{demons,demons2}. In addition, it can be used as a new resource for
quantum information processing protocols
~\cite{discordyingyong2,Luoshunlong,merging2},
and be harnessed as a measurable quantity to implement
quantum-enhanced tasks in the absence of quantum
entanglement~\cite{discordyingyong1,experiment1,experiment2}.

Majorana fermions (MFs) as quasi-particle excitations were
anticipated to occur in different solid-state systems, 
such as the $5/2$ fractional quantum Hall system
~\cite{nr}, the $p$-wave superconductor~\cite{ay}, and the hybrid systems of a
topological insulator~\cite{fk}. Recently, semiconductor nanowires
with strong spin-orbit coupling are attracting increasing interest.
In addition to the application in spin-orbit
qubit~\cite{lirui2,lirui}, these semiconductor nanowires provide a
platform for demonstrating MFs
~\cite{semiSarma1,nw1,Kouwenhoven,Shtrikman}.
Theoretically, it is shown that MFs can occur at the ends of a
semiconductor nanowire with strong spin-orbit coupling when the
nanowire is placed in proximity to an $s$-wave
superconductor
~\cite{semiSarma1,nw1}. It is expected that the
nanowire-superconductor system based on conventional
materials is more accessible to experimentally detecting the
MFs
~\cite{Kouwenhoven,Shtrikman}. In solid-state
systems, MFs are coherent superpositions of electron and hole
excitations. Two spatially separated MFs can form one fermionic
level, which can be either occupied or empty, and thus defines a
nonlocal qubit.
The paired MFs may exhibit intrinsic
nonlocal properties, so the search for the quantum effects induced
by such nonlocal paired MFs can provide an important signature for
the existence of MFs. Several schemes have been proposed to test the non-locality of MFs by studying the correlation behaviors of electron tunneling
through individual MFs~\cite{lilaoshi2,cebudao2,Bolech}. However, for
the completely separated MFs, the tunneling events through each MF are shown to be fully independent~\cite{lilaoshi2,cebudao2,Bolech}. Therefore, the
intrinsic non-locality of the paired MFs cannot be demonstrated in such a context.

A hybrid quantum system combining two or more physical systems
can often utilize the strengths of different systems to better
explore new phenomena and potentially bring about novel
quantum technologies~\cite{RMP}. For the two MFs emerging at the ends
of a semiconductor nanowire, because their energies are zero, it is
difficult to directly demonstrate their intrinsic non-locality,
particularly when the tunnel coupling between them becomes zero.
Thus, to probe the  emerged two MFs in the nanowire, in this work, we consider a hybrid quantum system consisting of two quantum dots (QDs) interacting with these two MFs.
To characterize the quantum correlations between the two QDs mediated by the paired MFs, we use both quantum entanglement and quantum discord. We show that when the paired MFs are
tunnel-coupled via the wavefunction overlapping, the maximally
entangled states can be induced from an unentangled (i.e., product)
state at specific dynamical time points. However, when the paired MFs are completely separated (i.e., no tunneling between them), quantum entanglement cannot be induced from this product state, but remarkably,
quantum discord is shown to still persist owing to the intrinsic
nonlocal correlation of MFs. This indicates that the quantum discord
can indeed reflect the non-locality of the paired MFs. Furthermore, we
propose an experimentally accessible approach to measuring the onset of quantum discord.
Our protocol provides a new method to probe the MFs emerging at the two ends of the nanowire.

\vspace{.8cm}
\noindent{\large\bf Results}

\vspace{.1cm}
\noindent
{\bf Quantum dynamics.~}We study a hybrid structure shown in Fig.~1, where a
semiconductor nanowire with strong spin-orbit coupling is placed in
proximity to an \emph{s}-wave superconductor. This nanowire is predicted to be driven into 
a topological superconducting phase
~\cite{semiSarma1,nw1}. A pair
of zero-energy MFs,  $\gamma_{1}$ and $\gamma_{2}$, are anticipated
to appear at the two ends of the nanowire. The Hamiltonian of the
paired MFs is described by $H_{M} =i
\epsilon_{m}\gamma_{1}\gamma_{2}/2$, where $\epsilon_{m}\sim
e^{-L/\xi}$ describes the tunnel coupling between the two MFs due to
the overlap of wavefunctions. This inter-MF coupling damps
exponentially with $L$ (the length of the nanowire) and $\xi$ (the
superconducting coherent length). Because the paired MFs have zero
energies, it is difficult to directly demonstrate their quantum
correlations, particularly when the tunnel coupling $\epsilon_{m}$
becomes zero. Thus, we introduce two QDs (i.e., $\textrm{QD}_{1}$ and 
$\textrm{QD}_{2}$) on both sides of the nanowire, each of which is tunnel-coupled to a MF neighboring to it.
In addition, we consider the strong Coulomb blockade regime for each
QD, so that only one electron is allowed therein when the tunneling
process takes place. The Hamiltonian of the two spatially separated
quantum dots is written as
$H_{DD}=\sum_{j=1,2}\varepsilon_{j}\emph{d}_{j}^{\dag}
\emph{d}_{j}$, where $\emph{d}_{j}^{\dag}$ ($\emph{d}_{j}$) is the
electron creation (annihilation) operator of the dot $j$, with the
corresponding energy $\varepsilon_{j}$. The tunneling Hamiltonian
between the QDs and the MFs is described by~\cite{Bolech,cebudao2,lilaoshi2}
$H_{TM}=\sum_{j=1,2}\lambda_{j}(\emph{d}_{j}^{\dag}
-\emph{d}_{j})\gamma_{j}$,
 with $\lambda_{j}$ $(j=1,2)$ denoting the tunnel coupling of the $j$th
 dot to its neighboring MF.

\begin{figure}
\begin{center}
\includegraphics[width=8.5cm]{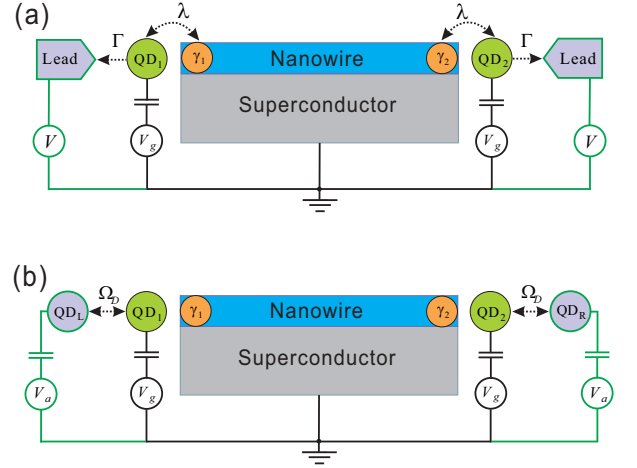}
\end{center}\caption{\textbf{Schematic setup for
probing quantum correlations between two QDs mediated by a
pair of MFs.} A semiconductor nanowire with strong spin-orbit
coupling is placed in proximity to an \emph{s}-wave superconductor,
with a pair of MFs emerging at the two ends of the nanowire. (a) Two electrode
leads are introduced to investigate the decoherence effects on the
quantum correlations.
(b) Two auxiliary QDs (denoted by $\textrm{QD}_{L}$
and $\textrm{QD}_{R}$) are introduced to measure the
quantum correlations. The arrows indicate the tunneling
directions of electrons. The energy levels of the QDs and the
chemical potentials of the electrode leads are tuned by the voltages
applied on them.}
\label{fig1}
\end{figure}

The total Hamiltonian of the system is
a sum of the above three parts: $H_{\textrm{sys}}
=H_{M}+H_{DD}+H_{TM}$. For practical calculations,  it is more convenient to switch
from the Majorana representation to the regular fermion one via the
transformations: $\gamma_{1}=i(f-f^{\dag})$, and
$\gamma_{2}=f+f^{\dag}$, where $f$ is the regular fermion operator,
satisfying the anticommutation relation $\{f,f^{\dag}\}=1$. After an
additional local gauge transformation $d_{1}\rightarrow i d_{1}$,
the total Hamiltonian of the system can be written in the new
representation as~\cite{Bolech,cebudao2,lilaoshi2}
\begin{eqnarray}\label{H-sys}
H_{\rm{sys}}
\!&\!=\!&\!\epsilon_{\emph{m}}(f^{\dag}f-1/2)+\sum_{j=1,2}[\varepsilon_{j}d_{j}^{\dag}d_{j}
+\lambda_{j}(\emph{d}_{j}^{\dag}f+f^{\dag} d_{j})]
\nonumber\\
&&\!-\lambda_{1}(d_{1}^{\dag}f^{\dag}\
+fd_{1})+\lambda_{2}(d_{2}^{\dag}f^{\dag}+fd_{2}).~~~~~~~~~~~~~
\label{Eq1}
\end{eqnarray}
For simplicity, we adopt a symmetric setup with
$\lambda_{1}=\lambda_{2}=\lambda$. Moreover, the two QDs are
adjusted in resonance with the two MFs (i.e.,
$\varepsilon_{1}=\varepsilon_{2}=0$) via the gate voltages. Also, we
use $\lambda$ $(1/\lambda)$ as the energy (time) unit throughout the
paper.

The QD-MF system considered in this paper includes the two QDs (i.e., $\textrm{QD}_{1}$ and $\textrm{QD}_{2}$) and the MFs. For the new fermion representation, it is convenient to use the state basis $|n_{1},n_{2},n_{M}\rangle$, where
$n_{l}=0$ or $1$, with $l=1,2,M$, denoting the
electron occupation number in the left $\textrm{QD}$ (i.e.,
$\textrm{QD}_{1}$ ), the right $\textrm{QD}$ (i.e., $\textrm{QD}_{2}$) and
the paired MFs, respectively. From Hamiltonian (\ref{H-sys}), one
may notice that the electron numbers in both QDs and the paired MFs
are not conserved, because a pair of electrons can be
extracted out from the superconductor or absorbed by it. As a consequence,
the parity of the isolated QD-MF system becomes conserved, and
the Hilbert space can be split into two subspaces: $|1,1,1\rangle$,
$|1,0,0\rangle$, $|0,1,0\rangle$, and $|0,0,1\rangle$ with odd
parity; $|1,1,0\rangle$, $|1,0,1\rangle$, $|0,1,1\rangle$,
and $|0,0,0\rangle$ with even parity. Thus, starting
from a product state $|0,0,1\rangle$, only odd-parity states are
involved in the state evolution:
\begin{eqnarray}\label{H-tms1y} 
|\Psi(t)\rangle
\!&\!=\!&\!C_{1}(t)|1,1,1\rangle+C_{2}(t)|1,0,0\rangle \nonumber\\
&&\!+C_{3}(t)|0,1,0\rangle
+C_{4}(t)|0,0,1\rangle.\end{eqnarray} 
With this initial condition, the time-dependent coefficients can be obtained, via directly
solving Schr\"odinger equation, as follows:
\begin{eqnarray}\label{xishu}
C_{1}(t)\!&\!=\!&\!\frac{-\Delta e^{-\emph{i}
\Omega t}+ \Delta
\cos(\Delta
t)-\emph{i}\Omega \sin(\Delta
t)}{2\Delta},\nonumber\\
C_{2}(t)\!&\!=\!&\!-\frac{\emph{i}\lambda\sin(\Delta
t)}{\Delta},~~
C_{3}(t)=-\frac{\emph{i} \lambda \sin(\Delta
t)}{\Delta},\\
C_{4}(t)\!&\!=\!&\!\frac{\Delta e^{-\emph{i}\Omega
t}+
\Delta\cos(\Delta t)-\emph{i}\Omega \sin(\Delta
t)}{2\Delta},~~\nonumber
\end{eqnarray}
where
$\Delta=\sqrt{\epsilon_{m}^{2}+16\lambda^{2}}/2$,
and
$\Omega=\epsilon_{m}/2$. From equation
(\ref{H-tms1y}), we can easily obtain the density operator
$\rho(t)=|\Psi(t)\rangle\langle\Psi(t)|$. After
tracing over the degrees of freedom of the MFs, the reduced density
operator $\rho_{d}(t)$ for the two QDs is obtained. This
reduced density operator contains the nonlocal information of
the paired MFs and serves as the starting point for calculating the quantum correlations (see Methods).

In order to investigate the decoherence effect on
quantum correlations, we also introduce two electrode
leads, each weakly coupled to its neighboring QD. The two
electrode leads can be treated as electron reservoirs described by the
Hamiltonian
$H_{\textrm{leads}}=\sum_{j=1,2}\sum_{k}\varepsilon_{kj}
\emph{c}_{k
j}^{\dag}c_{kj}$, where $c_{kj}^{\dag}$ ($c_{kj}$)
is the creation (annihilation) operator for the electron with wave
vector $k$ in the $j$th electrode lead and $\varepsilon_{kj}$ is the
corresponding energy. The tunneling Hamiltonian between the
electrode leads and
the QDs is $H_{TL}=\sum_{j=1,2}\sum_{k}[t_{j}
\emph{d}_{j}^{\dag}c_{kj}+\textrm{h.c.}]$. Here we
employ the master
equation approach to solving this problem. For the
whole setup, the leads are regarded as an environment, and the central QD-MF device is described by a
 reduced density operator $\rho(t)$. We assume that the chemical potentials of the two leads, both denoted by $\mu$, are well below the energy levels
 of the dots, i.e., the energy difference $\varepsilon_{1}-\mu$
 and $\varepsilon_{2}-\mu$ are much larger than the
 level broadening of the QDs. In this case, electrons can
only jump outwards from the central device to the electrodes.
In addition, we consider the regime with a weak coupling between
the QDs and the leads. Thus, we can use  the standard Born-Markov
master equation~\cite{sunhebi}. In the present case, the
master equation can be written as
\begin{equation}\label{rhoB}
 \dot{\rho}(t) =
-i[H_{\textrm{sys}},\rho]+\Gamma({\cal
D}[d_{1}]\rho(t)+{\cal
D}[d_{2}]\rho(t)),
\end{equation}
where ${\cal D}[A]\rho\equiv A\rho
A^{\dag}-\frac{1}{2}\{A^{\dag}A,\rho\}$, and the tunneling rate $\Gamma$ is associated with both the tunnel-coupling strengths and the densities of
states of the two electrode leads. With the density operator
$\rho(t)$ solved from equation (\ref{rhoB}), we can then obtain the
reduced density operator $\rho_{d}(t)$ of the two QDs by tracing
over the degrees of freedom of the MFs.

\vspace{.5cm}
\noindent
{\bf Quantum discord versus quantum entanglement.~}To measure quantum correlations, here we use both
quantum entanglement and quantum discord (see Methods). We
 show the evolutions of quantum entanglement and quantum
discord for the two QDs (i.e., $\textrm{QD}_{1}$ and $\textrm{QD}_{2}$) in Fig.~2.
As a typical example, we choose
 $\epsilon_{m} =0.5\lambda$ in Fig.~2a. In this
 case, the two MFs are tunnel-coupled, in addition to the intrinsic
nonlocal correlation of the paired MFs. Equation
(\ref{xishu}) shows that there are two frequencies determining the
oscillating behaviors of the quantum state $|\Psi(t)\rangle$. The fast
 frequency $\Delta/2\pi$ determines the period of the oscillations.
From equations (\ref{H-tms1y}) and (\ref{xishu}),
it can be seen that the state of the system becomes $C_{1}(t)|1,1,1\rangle+C_{4}(t)|0,0,1\rangle$ at $t=n\pi/\Delta$ $(n=1,2,3\cdots)$, i.e., the state of the two QDs is just an entangled state $C_{1}(t)|1,1\rangle+C_{4}(t) |0,0\rangle$. As shown in Fig.~2a, the quantum entanglement peaks at $t=n\pi/\Delta$. The slow frequency $\Omega/2\pi$ modulates the height of oscillating peaks and determines the period of the modulating cycles. It can be seen that the quantum
discord exhibits similar  behaviors of oscillations except for the very
short initial and final periods of each cycle, where the quantum entanglement becomes close to zero. Also, as  shown in  Fig.~2a, the quantum entanglement  (discord) can  reach its maximal value  $C_{d} = 1$ ($D_{d} = 1$) at some specific  time points. When $\epsilon_{m} = 0$, which corresponds to the long-length limit of the nanowire, the two MFs are completely separated in space. In this case, there is no direct tunnel coupling between the two MFs, but their intrinsic nonlocal correlation still exists. As shown in Fig.~2b, the temporal behaviors of quantum entanglement and quantum discord are very different at $\epsilon_{m} = 0$. While the quantum
entanglement remains zero with time, the quantum discord
oscillates with a smaller amplitude. Previous studies have found it hard to demonstrate the intrinsic nonlocal correlation
between the two MFs in the electrical measurement signals,
including the electrical current and the current
noise~\cite{lilaoshi2,Bolech,cebudao2}. Thus, it will be important to select a physical quantity that can significantly reflect the non-locality of the MFs. Interestingly, our results reveal that the quantum discord can clearly serve as an indicator showing the intrinsic nonlocal correlation
of the MFs in contrast to the insensitive quantum entanglement considered
in our system.

\begin{figure}
\includegraphics[width=3.35in,
bbllx=48,bblly=40,bburx=369,bbury=264]{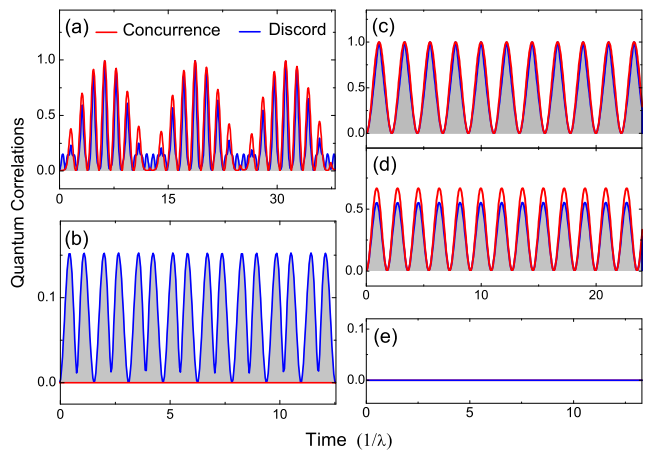} 
\caption{\textbf{Quantum dynamics of the
quantum correlations between two QDs mediated by different kinds of
fermions in the absence of quantum decoherence.}
 We assume a symmetric setup with
$\lambda_{\emph{1}}=\lambda_{\emph{2}}=\lambda$,
and $\varepsilon_{1}=\varepsilon_{2}=0$.
(a) Two QDs mediated by a pair of MFs with
an inter-MF coupling
$\epsilon_{m}=0.5\lambda$. (b) Two QDs mediated by a pair
of MFs without an inter-MF coupling (i.e., $\epsilon_{m}=0$). (c)
Two QDs mediated by a regular fermion with energy $\epsilon_c=0$. (d)
Two QDs mediated by a regular fermion with $\epsilon_c=\lambda$,
(e) Two QDs mediated by a pair of non-interacting regular fermions
with energies $\epsilon_1=\epsilon_2=0$.}
\label{fig2}
\end{figure}

For the purpose of discriminating MFs from regular fermions, we also show the quantum correlations of the two QDs (i.e., $\textrm{QD}_{1}$ and $\textrm{QD}_{2}$) induced by regular fermions that couple to them. We first consider
the case with the two QDs coupled by a common regular fermion. Different from equation (\ref{Eq1}), the Hamiltonian of this QD-fermion system is given by $H_{\rm{sys}}
=\epsilon_{c}c^{\dag}c+\sum_{j=1,2}[\varepsilon_{j}d_{j}^{\dag}d_{j}
+\lambda_{j}(\emph{d}_{j}^{\dag}c+c^{\dag}d_{j})]$, where $\emph{c}^{\dag}$ ($\emph{c}$) is the electron creation (annihilation) operator of the regular fermion with energy $\epsilon_{c}$, and $\lambda_{j}$ $(j=1,2)$ denotes the tunnel coupling between the $j$th dot and the regular fermion. We assume that the two QDs is initially empty, but the fermionic level $\epsilon_c$ is occupied by a single regular fermion such as the fermionic quasi-particle excitation in the nanowire.
When the fermionic level $\epsilon_c$ is resonant to the energy levels $\varepsilon_i$ of the two QDs, both quantum discord and quantum entanglement of the two QDs induced by the fermion have identical dynamical evolutions (see Fig.~2c); when $\epsilon_c$ is off-resonant to $\varepsilon_i$, these quantum discord and quantum entanglement have different amplitudes, but still exhibit very similar temporal behaviors. Moreover, we consider the case involving a pair of non-interacting regular fermions, where each fermion couples to a QD. The Hamiltonian of this QD-fermion system is
$H_{\rm{sys}}=\sum_{j=1,2}[\varepsilon_{j}d_{j}^{\dag}d_{j}
+\epsilon_{j}c_{j}^{\dag}c_{j}+\lambda_{j}(\emph{d}_{j}^{\dag}c_{j}+c_{j}^{\dag}
d_{j})]$, where $\emph{c}_{j}^{\dag}$ ($\emph{c}_{j}$) is the
electron creation (annihilation) operator of the regular fermion with
energy $\epsilon_{j}$, and $\lambda_{j}$
$(j=1,2)$ denotes the tunnel coupling between the $j$th quantum dot and the regular fermion that couples to it. Also, we assume that the two QDs are initially empty, but each fermionic level $\epsilon_i$ is occupied by a single regular fermion.
Because there is no intrinsic non-locality between these two non-interacting regular fermions, both quantum discord and quantum entanglement of the two QDs remains zero with time (see Fig.~2e). As compared with Figs.~2a and 2b, these very different temporal behaviors in Figs.~2c-2d can be used to discriminate the MFs from the regular fermions.

We also study the temporal behaviors of both
quantum entanglement and quantum discord for the QD-MF system in the
presence of two electrode leads. Here we choose a tunneling rate
$\Gamma=0.05\lambda$ to ensure a weak coupling
between the central QD-MF system and the two electrode leads. These two
leads play the role of a fermionic environment which induces
decoherence on the evolutions of quantum correlations. With tunnel
coupling between the paired MFs, e.g., $\epsilon_{m} =0.5\lambda$ in
Figs.~3a, both quantum entanglement and quantum discord can be
generated from the product state, but the oscillating quantum
entanglement decays faster with time than the quantum discord. This implies that the quantum discord, as a new resource of quantum information processing, is more robust than the quantum
entanglement in this model system. Moreover, without tunnel coupling between the paired MFs, i.e., $\epsilon_{m} = 0$, the quantum entanglement still remains zero with time, but the oscillating quantum discord decays due to the decoherence induced by the electric leads (see Fig.~3b).

\vspace{.5cm}
\noindent
{\bf A measurement scheme.~}Note that equation (\ref{H-tms1y}) can be rewritten as
\begin{eqnarray}\label{t0b}
|\Psi(t)\rangle
\!&\!=\!&\!b_{1}(t)e^{\emph{i}\phi_{1}(t)}|1,1,1\rangle+b_{2}(t)|1,0,0\rangle \nonumber\\
&&\!+b_{2}(t)|0,1,0\rangle
+b_{3}(t)e^{\emph{i}\phi_{3}(t)}|0,0,1\rangle,
\end{eqnarray}
where the real parameters $b_{1}(t)$, $b_{2}(t)$,
$b_{3}(t)$, $\phi_{1}(t)$ and $\phi_{3}(t)$ are determined by
$C_{i}(t)$, $i=1, 2, 3,$ and $4$, in equation (\ref{xishu}). After
tracing over the degrees of freedom of the MFs, the reduced density
operator of the two QDs (i.e., $\textrm{QD}_{1}$ and
$\textrm{QD}_{2}$) can be obtained via equation (\ref{t0b}) as
\begin{eqnarray}\label{juzhen1}
\rho_{d}(t)\!&\!=\!&\! \left(\!\!
\begin{array}{cccc}
 b_1^2(t) \!\!\!&\! 0\! &\! 0\! &\!\!\! b_1(t) b_3(t) e^{i \Delta
 \phi(t)} \\
 0\!\!\! &\! b_2^2(t) \!&\! b_2^2(t)\! &\!\!\! 0 \\
 0\!\!\!&\! b_2^2(t)\! &\! b_2^2(t)\! &\!\!\! 0 \\
 b_1(t) b_3(t) e^{-i \Delta \phi (t)} \!\!\!&\! 0 \!&\! 0\! &\!\!\!
 b_3^2(t)
\end{array}
\!\!\right),\nonumber\\
&&
\end{eqnarray}
with $\Delta \phi(t)=\phi_{1}(t)-\phi_{3}(t)$. Here
$b_1^2(t)$, $b_2^2(t)$ and $b_3^2(t)$, i.e., the diagonal
matrix elements in equation (\ref{juzhen1}), correspond to the
probabilities of both dots occupied, either dot occupied, and both dots
empty, respectively. These probabilities can be directly
obtained by joint measurements on the electron occupation of
$\textrm{QD}_{1}$ and $\textrm{QD}_{2}$ (see, e.g., Ref.~~\cite{moc1}).
In order to obtain the values of the phase difference
$\Delta \phi(t)$, we introduce, as Fig.~1b shows, two auxiliary QDs (denoted by $\textrm{QD}_{L}$ and $\textrm{QD}_{R}$) adjacent to $\textrm{QD}_{1}$ and
$\textrm{QD}_{2}$, respectively. First, instead of $|0\rangle$,
these two auxiliary QDs are initially prepared in the superposition
state: $|\psi_{L(R)}\rangle=(|0\rangle_{L(R)}+|1\rangle_{L(R)})/\sqrt{2}$.
Second, the $\textrm{QD}_{L}$ ($\textrm{QD}_{R}$) is adjusted in
resonance with the $\textrm{QD}_{1}$ ($\textrm{QD}_{2}$) (i.e.,
$\varepsilon_{L}=\varepsilon_{1}=0$, and $\varepsilon_{R}=\varepsilon_{2}=0$) at the moment $t$, via tuning the gate voltages. Simultaneously, the tunnel coupling between $\textrm{QD}_{1}$ ($\textrm{QD}_{2}$) and its
adjoining MF is switched off by tuning the voltages of the
electrical gates that control the tunnel barrier between them.  The
Hamiltonian of the total system becomes $H_{TD}
=T_{1}(\emph{d}_{L}^{\dag}d_{1}+d_{1}^{\dag}d_{L})+
T_{2}(d_{R}^{\dag}d_{2}+d_{2}^{\dag}d_{R})$
, where $T_{1}$ ($T_{2}$) is the tunnel coupling between
$\textrm{QD}_{L}$ ($\textrm{QD}_{R}$) and
$\textrm{QD}_{1}$ ($\textrm{QD}_{2}$) and assume $T_{1}=T_{2}=T$.
Therefore, the system further
evolves under the Hamiltonian $H_{TD}$, from the initial product
state $|\psi_{L}\rangle\otimes|\psi_{R}\rangle\otimes|\Psi(t)\rangle$.
Third, along with the evolution of the total
system, we choose a specific moment, $t+\triangle t$, to jointly
measure the occupation probabilities of both $\textrm{QD}_{1}$ and
$\textrm{QD}_{2}$, i.e., $\textrm{P}_{11}$, $\textrm{P}_{10}$,
$\textrm{P}_{01}$, and $\textrm{P}_{00}$ which denote the probabilities of
both dots occupied, single $\textrm{QD}_{1}$ occupied, single
$\textrm{QD}_{2}$ occupied, and both dots empty, respectively.
 When $\triangle t=\pi/4T$, these probabilities
can be obtained as 
\begin{eqnarray}\label{xishuy1}
\textrm{P}_{11}(t)\!&\!=\!&\!\frac{1}{16} \left[9
b_1^2 (t)+8 b_2^2(t)+b_3^2(t)\right. \nonumber\\
&&\left.~~~~~~~~~~~~~~~~~~~~~~~+2 b_1(t) b_3(t)
\cos \Delta \phi (t)\right],\nonumber\\
 \textrm{P}_{00}(t)\!&\!=\!&\!
\frac{1}{16} \left[b_1^2(t)+8
b_2^2 (t)+9 b_3^2(t)\right. \nonumber\\
&&\left.~~~~~~~~~~~~~~~~~~~~~~+2 b_1(t) b_3(t) \cos\Delta
\phi (t)\right],~~~~\\
 \textrm{P}_{10}(t)\!&\!=\!&\!\frac{1}{16} \left[3
b_1^2(t)+8 b_2^2(t)+3 b_3^2(t)\right. \nonumber\\
&&\left.~~~~~~~~~~~~~~~~~~~~~~~-2 b_1(t) b_3(t) \cos
\Delta \phi (t)\right],\nonumber 
\end{eqnarray}
and $\textrm{P}_{01}(t)=\textrm{P}_{10}(t)$. This
measurable occupation probabilities include the phase difference $\Delta
\phi (t)$, in addition to $b_1(t)$, $b_2(t)$, and $b_3(t)$ that
have previously been obtained via joint measurements as well. Thus,
it is experimentally feasible to obtain  the reduced
density operator $\rho_{d}(t)$ of the two QDs mediated by the paired
MFs. Once the reduced density operator $\rho_{d}(t)$ is
determined experimentally, one can then  derive the quantum entanglement and
the quantum discord of the two QDs using this reduced density
operator (see Methods).

\begin{figure}
\begin{center}
\includegraphics[width=7.5cm]{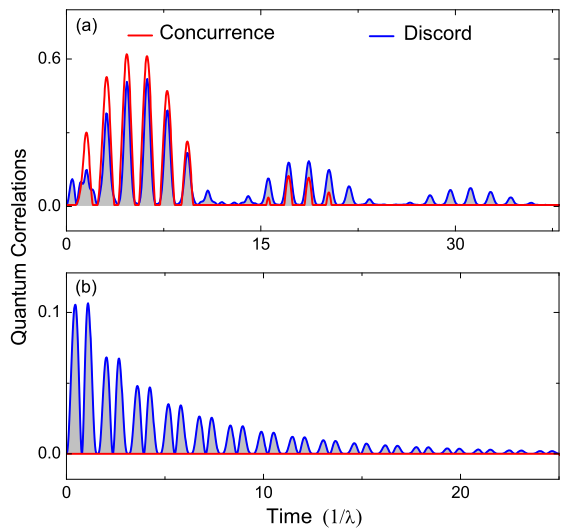}
\caption{\label{fig2}\textbf{Quantum dynamics of the
quantum correlations between two QDs mediated by a pair of
MFs in the presence of quantum decoherence.}
We assume a symmetric setup with
$\lambda_{\emph{1}}=\lambda_{\emph{2}}=\lambda$,
and
$\varepsilon_{1}=\varepsilon_{2}=0$. The tunneling
rate between the central QD-MF device and the two electrode leads is
chosen as $\Gamma=0.05\lambda$.
(a) $\epsilon_{m}=0.5\lambda$,
and (b) $\epsilon_{m}=0$.}
\end{center}
\end{figure}

\vspace{.5cm}
\noindent
{\large\bf Discussion}

\vspace{.1cm}
\noindent
We have investigated the dynamics of quantum correlations between
two QDs mediated by a pair of MFs. We find that in the presence of
tunnel coupling between the two MFs, maximally entangled states of
the two QDs can be generated from
an unentangled state under proper dynamical conditions. Notably, in the absence of tunnel coupling
between MFs, while the quantum entanglement remains zero,
quantum discord can be induced from this unentangled state owing to
the intrinsic nonlocal correlations of the MFs. While the conventional methods fail to reveal the non-locality information of the MFs, our demonstrated new feature indicates that the quantum discord can serves as a novel quantum signature
reflecting the intrinsic non-locality of the MFs.  Physically, it shows that quantum discord captures some interesting correlation features of the system that is insensitive to entanglement. Also, we show that these features do not exist when replacing MFs with regular fermions. Therefore, they can be used to discriminate the MFs from the regular fermions. Furthermore, we have proposed a scheme for experimentally determining the quantum discord of the two QDs mediated by the paired MFs. In addition, we further investigate the decoherence effects on the quantum correlations, and the results show that the quantum discord is more
robust than the quantum entanglement. Because the quantum discord is
a measurable physical quantity and the intrinsic nonlocal nature is
an extraordinary character of the MFs, our approach provides a new,
and experimentally feasible, method to probe the MFs by
demonstrating their intrinsic non-locality via the quantum discord.

\vspace{.5cm}
\noindent
{\large\bf Methods}

\vspace{.1cm}
\noindent
With the reduced density operator $\rho_{d}$ of the two QDs (i.e., $\textrm{QD}_{1}$ and $\textrm{QD}_{2}$) , the
concurrence~\cite{concurrence}, as a measure of the quantum
entanglement, can be obtained as \begin{equation}\label{concurrence}
C_{d}=\textrm{max}\{0,\sqrt{\Lambda_{1}}-\sqrt{\Lambda_{2}}-
\sqrt{\Lambda_{3}}-\sqrt{\Lambda_{4}}\},\end{equation}
where $\Lambda_{1}$, $\Lambda_{2}$, $\Lambda_{3}$ and
$\Lambda_{4}$ are the square roots of the eigenvalues in decreasing
order for the matrix \begin{equation}\label{rhoa}
G=\rho_{d}(\sigma_{y}\otimes\sigma_{y})\rho_{d}^{*}
(\sigma_{y}\otimes\sigma_{y}),\end{equation} where
$\sigma_{y}$ is a Pauli matrix, and $\rho_{d}^{*}$ is the complex
conjugate of $\rho_{d}$.

Quantum discord is a measure of quantum correlations obtained by
subtracting the classical correlations from the total correlations~\cite{discordPRL,discordVedral}. For a bipartite quantum
system including subsystems $\emph{A}$ and $\emph{B}$, the total
correlations are defined, in terms of quantum mutual information, as
$\emph{I}(\rho^{A}\!:\!\rho^{B})=\emph{S}(\rho^{A})+\emph{S}(\rho^{B})
-\emph{S}(\rho_{AB})$, where $\rho_{AB}$ denotes the density
operator of the bipartite system, with $\rho^{A}$ ($\rho^{B}$) being
the reduced density operator of the subsystem $\emph{A}$ ($\emph{B}$), and
$S(\rho_{AB})=-\textrm{Tr}(\rho_{AB} \textrm{log}_{2}\rho_{AB})$ is the entropy of the bipartite quantum system. The total classical correlations based on the POVM measurement of the subsystem $\emph{B}$ are defined as
~\cite{discordVedral,Vedral2}
$\emph{J}(\rho_{AB}\!\mid\!\{\emph{B}_{\emph{k}}\})=
\emph{S}(\rho^{A})-\textrm{min}_{\{\emph{B}_{\emph{k}}\}}\sum_{\emph{k}}\emph{p}_{\emph{k}}
\emph{S}(\rho_{\emph{k}})$, where $\{\emph{B}_{\emph{k}}\}$ is a
complete set of projectors of local measurements performed on the
subsystem $B$; $\emph{p}_{\emph{k}}=\textrm{Tr}[(\emph{I}\otimes
\emph{B}_{\emph{k}})\rho_{AB}(\emph{I} \otimes
\emph{B}_{\emph{k}})]$ denotes the probability of
obtaining the measurement outcome $\emph{k}$, with $\emph{I}$
being the identity operator for the subsystem $\emph{A}$;
$\rho^{\emph{k}}=(\emph{I} \otimes \emph{B}_{\emph{k}})\rho_{AB}(\emph{I}
\otimes \emph{B}_{\emph{k}})/\emph{p}_{\emph{k}}$ is the
state of the bipartite system, conditioned on the measurement
outcome labeled by $\emph{k}$. The quantum discord is defined
as~\cite{discordPRL,discordVedral}
\begin{equation}\label{discordy}
\emph{D}_{B}(\rho_{AB})=\textrm{min}_{\{\emph{B}_{\emph{k}}\}}
[\emph{I}(\rho^{A}:\rho^{B})-\emph{J}(\rho_{AB}\mid\{\emph{B}_{\emph{k}}\})].
\end{equation}
For our system, the $A$ and $B$ parts are the two QDs (i.e., $\textrm{QD}_{1}$ and $\textrm{QD}_{2}$) , and $\rho_{AB}=\rho_{d}$. 
To compute classical correlations, we adopt the
specific case of von Neumann local measurements, which are provided by
the orthogonal projectors
$B_{\emph{k}}=R\prod_{\emph{k}}R^{\dag}$, where
$\{\prod_{\emph{k}}\}=\{|\emph{k}\rangle\langle\emph{k}|\}$
defines the computational basis $R$, with $\emph{k}=0, 1$
corresponding to the empty and occupied states, respectively. As a
rotation operator, $V$ can be parameterized as
\begin{eqnarray}\label{discord} R=\left(
    \begin{array}{cc}
      \textrm{cos}\frac{\theta}{2} &
      \textrm{sin}\frac{\theta}{2}e^{-\emph{i}\phi}
      \\
      -\textrm{sin}\frac{\theta}{2}e^{\emph{i}\phi}
      & -\textrm{cos}\frac{\theta}{2} \\
    \end{array}
  \right),\end{eqnarray}
where $\theta\in[0,\pi]$, and $\phi\in[0,2\pi]$.
Hence, the quantum discord  given in equation (\ref{discordy}) can be
directly obtained by minimizing over all angles $\theta$ and $\phi$.

\vspace{.5cm}
\noindent
{\bf Acknowledgement}

\vspace{.01cm}
\noindent
J.L. and J.Q.Y. are partially supported by the National Natural
Science Foundation of China Grant Nos.~91121015 and 11304031, the
National Basic Research Program of China Grant No.~2014CB921401, and
the NSAF Grant No.~U1330201.
T.Y. is partially supported by the NSF PHY-0925174 and AFOSR No.~FA9550-12-1-0001.
H.Q.L. is partially supported by the National Natural Science Foundation of China Grant No.~91230203, the National Basic Research Program
of China Grant No.~2011CB922200 and the NSAF Grant No.~U1230202.




\end{document}